\documentclass[aps,prb,showpacs,twocolumn]{revtex4}
\usepackage{amssymb}
\usepackage{amsmath}
\usepackage{graphicx}
\usepackage{amsfonts}
\usepackage{subfigure}
\usepackage{textcomp}
\usepackage{color}
\usepackage{amsfonts}
\usepackage{epsfig}
\usepackage{hyperref}

\def\beq{\nopagebreak \begin{equation}}
\def\eeq{\end{equation}}
\def\id{{\textrm d}}

\newcommand{\arctanh}[1]{\operatorname{arctan}}

\begin{document}

\title{High-throughput exploration of alloying as design strategy for thermoelectrics.}

\author{Sandip Bhattacharya and Georg K.H. Madsen}

\affiliation{ICAMS, Ruhr-Universit\"at Bochum, 44780 Bochum, Germany }

\date{\today}

\begin{abstract}
\vspace{0.5cm}

We explore a material design strategy to optimize the thermoelectric power factor. The approach is based on screening the band structure changes upon a controlled volume change. The methodology is applied to the binary silicides and germanides. We first confirm the effect in antifluorite Mg$_2$Si and Mg$_2$Ge where an increased power factor by alloying with Mg$_2$Sn is experimentally established. Within a high-throughput formalism we identify six previously unreported binaries that exhibit an improvement in their transport properties with volume. Among these, hexagonal MoSi$_2$ and orthorhombic Ca$_2$Si and Ca$_2$Ge have the highest increment in $zT$ with volume. We then perform super-cell calculations on special quasi-random structures to investigate the possibility of obtaining thermodynamically stable alloy systems which would produce the necessary volume changes. We find that for Ca$_2$Si and Ca$_2$Ge the solid solutions with the isostructural Ca$_2$Sn readily forms even at low temperatures.

%We will explore in detail the correlation between the electronic structures and the thermoeletric properties in these candidates to understand the source of the improvement in their transport properties. Finally, we will also explore the possibility to obtain thermodynamically stable alloy systems which would produce the volume changes necessary.

\end{abstract}

% refs: morelli
\maketitle
Despite their importance, the discovery of new materials are often based on trial and error. High-throughput (HT) computational screening\cite{Curtarolo,HT_exp} is an important step towards identifying materials with desired properties in a more systematic way. Thermoelectric (TE) materials are attractive for such computational searches because continuous development of computational methodology means that all parts of the TE figure of merit, $zT$ can in principle be calculated from first principles.\cite{davidskut2,Broido_APL07,Restrepo_APL09,GChen_EPL2015} In practice computational HT searches for new TE materials have focused on parts of the $zT=S^2\sigma T/\kappa$, where $S$ is the Seebeck coefficient, $\sigma$ the electrical and $\kappa$ the thermal conductivity.\cite{gmLiZnSb,Yang_AFM08,Wang_PRX11,Opahle_NJP13,Carrete_PRX14} Despite this, there are now a few works where computational screening has led to high performance TE materials that could be experimentally realized.\cite{chandan,sandip,Tan_JMCA14,Joshi_EES14} 

\begin{figure}
\includegraphics[width=.35\textwidth]{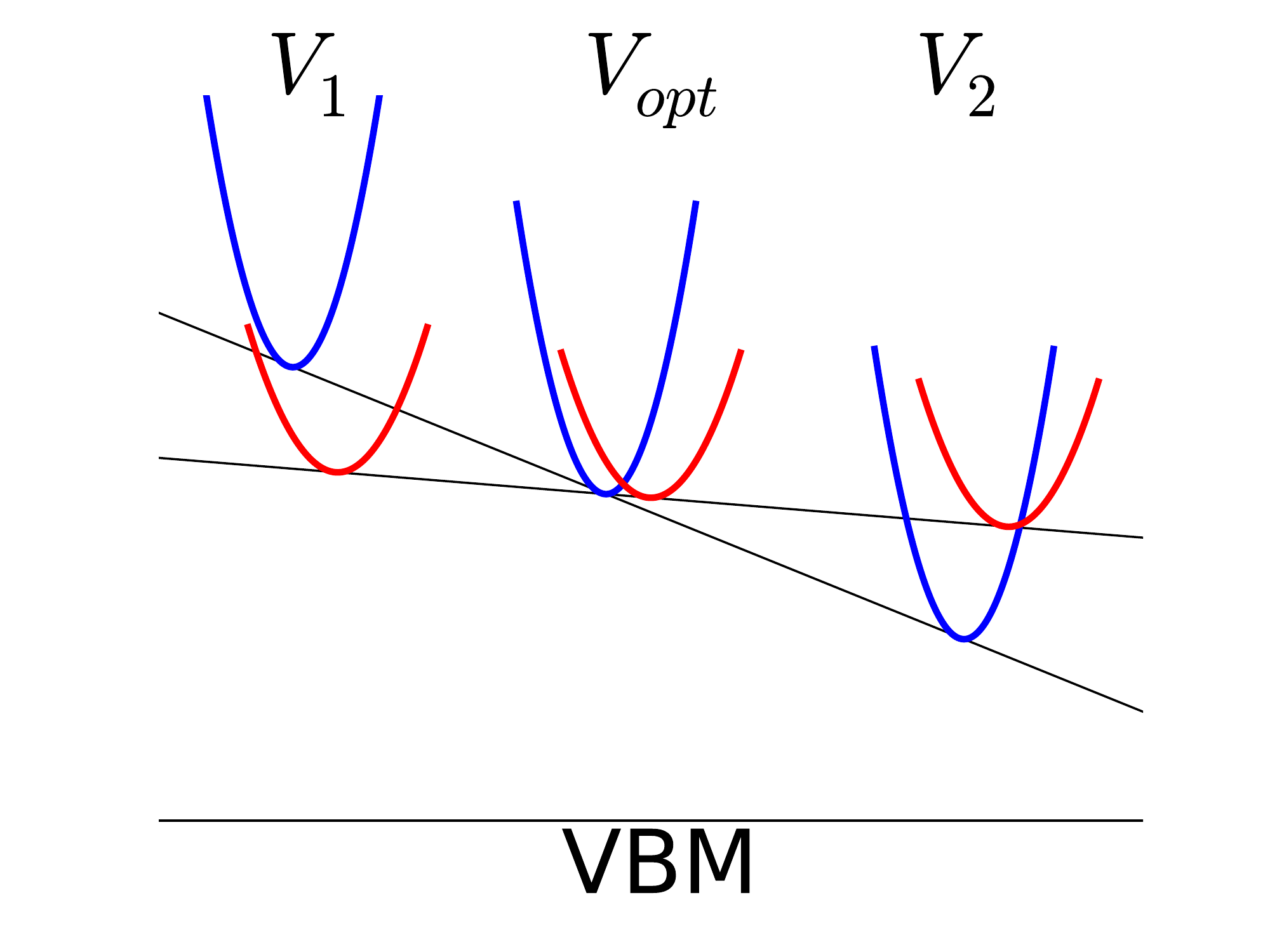}
\caption{Schematic illustration of volumetric band alignment for a $n$-type material. VBM stands for the Valence Band Minimum.}
\label{fig:vba}
\end{figure}
Beyond screening known compounds, there still exists great challenges designing new materials with specific properties. This is especially the case for electronic structure dependent properties, which have highly non-trivial dependencies on the atomic structure.\cite{Yan_EES14}
For TE materials one strategy for designing new alloys with optimized properties is by a controlled volume change. We label this procedure volumetric band-structure alignment (VBA). The idea is illustrated in Fig.~\ref{fig:vba}, where the energy dependence of two bands vary differently upon a change of volume. Thereby a scenario can occur when the band edges are aligned, as schematically illustrated in the mid-panel of Fig.~\ref{fig:vba}. How this optimizes the TE power factor, $PF=S^2\sigma$, can be understood by considering the generalized transport coefficients,
\beq
{\mathcal L}^{(\alpha)}=q^2\int \sigma(\varepsilon) (\varepsilon-\mu)^\alpha\biggl(-\frac{\partial f}{\partial \varepsilon}\biggr)\id\varepsilon,
\label{eq:Ltrans}
\eeq
where $f$ is the Fermi-distribution and $\sigma(\varepsilon)$ the transport distribution. The Seebeck and electric conductivity are given as $S={\mathcal L}^{(1)}/qT{\mathcal L}^{(0)}$ and $\sigma={\mathcal L}^{(0)}$, respectively. For two channels (labelled $'$ and $''$) conducting in parallel, ${\mathcal L}^{(\alpha)}$ is given as the sum of the contributions from each channel, so that the PF is,
\begin{equation}
S^2\sigma=\frac{1}{q^2T^2}\frac{\bigl({\mathcal L}^{(1)'}+{\mathcal L}^{(1)''}\bigr)^2}{{\mathcal L}^{(0)'}+{\mathcal L}^{(0)''}}.
\label{eq:poweropt}
\end{equation}

Without loss of generality we can write ${\mathcal L}^{(1)''}=\alpha_1{\mathcal L}^{(1)'}$ and ${\mathcal L}^{(0)''}=\alpha_0{\mathcal L}^{(0)'}$. Thereby it is clear that the $PF$ will be increased if,
\beq
(1+\alpha_1)^2>1+\alpha_0.
\label{condition}
\eeq
${\mathcal L}^{(1)}$ and thereby $S$ will be significant when the chemical potential is located within the relatively narrow energy window defined by ${\partial f}/{\partial \varepsilon}$ of a band edge. Aligning band edges with similar effective masses within ${\partial f}/{\partial \varepsilon}$, will satisfy Eq.~\eqref{condition} and lead to an increased $PF$ when compared to the largest of the individual contributions. Such bandstructure characteristics are found in several materials with complex carrier pocket shapes which results in enhanced TE properties.\cite{Bjerg_CM11,Parker2013_nonparab,Mecholsky_PRB14,Parker2015_AgBiSe}
%
%Eqs.~\eqref{eq:Ltrans}-\eqref{eq:poweropt} mean that if the edges of two bands are aligned within the relative narrow energy range defined by ${\partial f}/{\partial \varepsilon}$, the $PF$ will be increased compared to the individual contributions.

VBA has been successfully applied to optimize the PF in the $n-$doped Mg$_2$Si$_{1-x}$Sn$_x$ (and Mg$_2$Ge$_{1-x}$Sn$_x$) and in $p-$doped PbTe$_{1-x}$Se$_x$ alloys.\cite{mg2si_paper,pbte_paper} In both the cases the volume was controlled by alloying. Consider as an example the alloy composition MX$_{1-x}$X'$_x$. According to Vegard's law, the volume of the resulting alloy system, will be directly proportional to the fraction of the new composition added, i.e. $x$. The volume will also be influenced by the thermal expansion, and this together with alloying offers the possibility to control the peak temperature of TE performance.\cite{Gibbs_APL13}

The question whether it is possible to optimize the band structure by VBA for a given compound, is very difficult to answer from intuition alone. It depends on the detailed band structure and how the different bands react to a volume change. The effect of VBA can be strongly affected by changes to the band gap and doping level. Furthermore, it can be very difficult to predict whether a given compound can be alloyed or not. E.g. while a solid solution of Si and Sn in Mg$_2$Si$_{1-x}$Sn$_{x}$ can form even at low temperatures\cite{Viennois2012,Jung2007192,Jung2010,Kozlov20113326} it is well known that Sn is hardly soluble in diamond-Si.

The idea behind the present work is to explore a HT computational strategy to identify systems where VBA can be applied. The approach is based on screening the volumetric effect on the band structure and calculate ab-initio thermodynamics to asses the possibility of alloying. It is shown that systems allowing VBA are quite rare and that the procedure leads to a very strong screening of potential candidates. We focus on the electronic part of the $zT$. However, alloying will also be advantageous in terms of reducing the thermal conductivity either by lowering mass-disorder scattering or by using natural solubility limits to nano-structure the material.\cite{Zhang_APL08}

We will explore binary alloys of the group 14 elements. Silicon is the second most abundant element on the earth's crust \cite{abundance} making such systems quite attractive. Stable alloys and solid solutions with optimized electronic structures and transport properties have been realized with many binary silicides plus the corresponding germanides or stannides compounds.\cite{mg2si_paper} While Ge and Sn are not as abundant and cheap as Si, they are certainly not rare elements either.\cite{price} Other crucial advantages that these M-Si/Ge/Sn alloy systems enjoy, is the ease with which they can be doped. Indeed silicide group compounds exhibit a high density of mobile charge carriers upon doping of up to $n=10^{20}-10^{21}$ cm$^{-3}$. Furthermore, they often have high melting points making them attractive stable candidates for high-$T$ TE.\cite{fedorov}

%In the next section we will introduce the computational details used to scan the thermodynamically stable structures for different binary systems. Furthermore, we will also explain in detail the HT formalism used in this paper, including a discussion on the descriptor that captures the volumetric effect. Thereafter, in the first part of the results section we shall identify the binary systems that exhibit a large value of the high-throughput descriptor. Subsequently in the second part of the results section, we will explore in detail the origin behind the volumetric effect in each of the identified binary systems. Finally, in the last part of this paper, we conclude our work, comment on the possible outcome on realization of of the identified candidates in future thermoelectrics and list the promising new candidates identified through this work.

\section{Methods}

\subsection{High-throughput scheme}

\begin{figure}
\includegraphics[width=.5\textwidth]{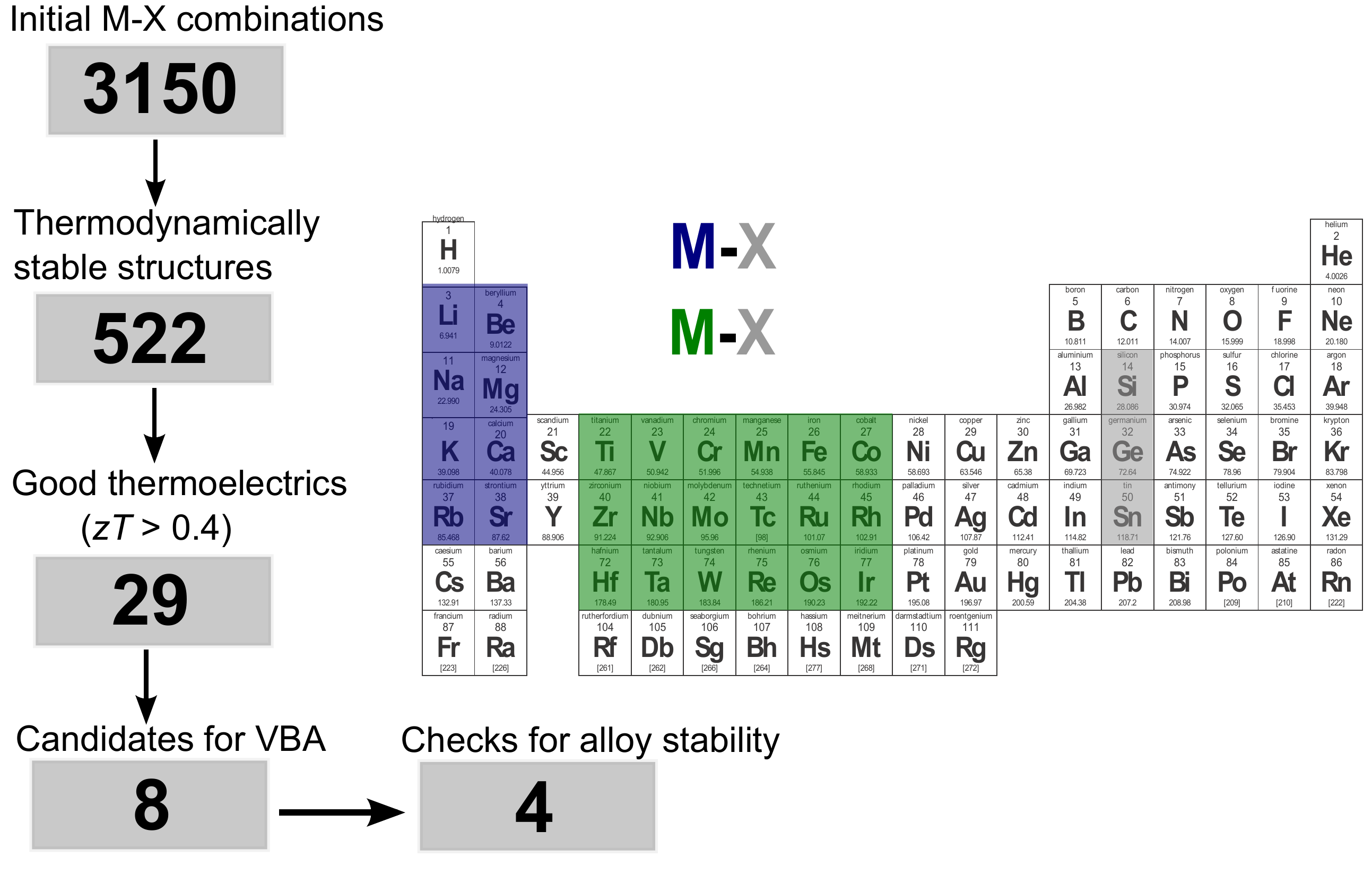}
\caption{Flowchart of the HT procedure employed to investigate the candidates for volumetric band engineering. The criterion for good thermoelectrics, i.e. $zT>0.4$, within our scheme is checked at $T=600$~K and $n=2\times10^{20}$~cm$^{-3}$ with $zT$ evaluated from Eq.~\eqref{zt_def}.}
\label{flowchart}
\end{figure}

Our HT scheme is illustrated in Fig.~\ref{flowchart}. We consider binary M-X systems, where M is a metal taken to be a $3d$, $4d$ or $5d$ Transition Metals (TM) or $2s$, $3s$ and $4s$ Alkali (A) or Alkaline Earth (AE) metals and X is Si, Ge or Sn. The first step of our work is to identify the stable structures in the constituent binary system. The phase stability is evaluated within a high throughput formalism based on our previous works.\cite{ingo2} For each metal silicide (or germanide) combination, the crystal structures were relaxed and the formation energy was calculated. The candidate structures were generated by the extracting the known structures from the Pearson's database\cite{pearsons} and appropriate substitution of the known structures with similar atoms (in same group of Periodic Table). Furthermore, we have also incorporated the silicides structures from our previous work\cite{ingo2} and consequent substitution of similar atoms.

In total, we have investigated 3150 different compounds. %Therefore, to facilitate the generation and analysis of the corresponding transport properties we need to build a HT machinery which adheres to the following scheme. 
For a given M-X binary system, we select the structures that have the difference in formation energies from the corresponding convex hull, $\Delta E_h$, lower than 50~meV/atom. The $\Delta E_h<50$~meV/atom tolerance has statistically been shown to contain 80~\% of the experimentally known compounds in the TM-Si system\cite{ingo2} and narrows the original number to 522 thermodynamically stable compounds, Fig.~\ref{flowchart}. 

Thereafter, the self-consistent calculations for these selected structures were performed using the (L)APW+lo method \cite{lapw} implemented within the WIEN2k code.\cite{wien2k}. These were followed by bandstructure calculations on a finer \textbf{k}-mesh of $\frac{64\times10^6}{V_{unitcell}}$ \textbf{k}-points in the full BZ. All calculations in this work are reported for Perdew-Burke-Ernzerhof (PBE) \cite{pbe} exchange-correlation potential. Subsequently, the electronic transport properties were evaluated using the BoltzTrap code.\cite{boltztrap} The code evaluates the Seebeck coefficient on an absolute scale and electrical conductivity in terms of the carrier relaxation time $\tau$ using the rigid band approximation. To evaluate the potential for TE energy conversion we utilize a modified version of the definition of figure of merit, $zT$,
\begin{equation}\label{zt_def}
zT=\frac{S^2 \frac{\sigma}{\tau}T}{L_o\frac{\sigma}{\tau}T+\frac{\kappa_{\mathrm{ph}}}{\tau}},
\end{equation}
The denominator in Eq.~\eqref{zt_def} accounts for the total thermal conductivity of the material, with its electronic part, $\kappa_{\mathrm{el}}=L_o \frac{\sigma}{\tau} T$, written according to the Wiedemann-Franz law. Here $L_o$ is the Lorenz number, $L_o= 2.44 \times 10^{-8}$ W$\Omega$K$^{-2}$. Finally the phononic part of thermal conductivity is $\kappa_{\mathrm{ph}}$.
All the transport quantities in Eq.~\eqref{zt_def} are extracted on an absolute scale, apart from $\frac{\kappa_{\mathrm{ph}}}{\tau}$ which is set at $10^{14}$ W K$^{-1}$ ms$^{-1}$.\cite{gmLiZnSb} In reality, the magnitude of $\frac{\kappa_{\mathrm{ph}}}{\tau}$ will be dictated by both intrinsic properties such as the phonon band structure and phonon- and electron-phonon coupling and by the details of sample preparation control e.g. grain size (nanostructuring) or the existence of multiple phases. $zT$, as defined in Eq.~\eqref{zt_def}, epitomizes the electronic contribution to the TE performance and has been validated in the prediction of TE materials.\cite{gmLiZnSb,Bjerg_CM11,Opahle_NJP13} It can be viewed as a descriptor for identifying potential high TE performance under the assumption of a low thermal conductivity, a long electron life time and the possibility of reaching the optimal doping. Setting a criteria of $zT>0.4$ [evaluated using Eq.~\eqref{zt_def}] at $T=600$~K and $2\times10^{20}$ cm$^{-3}$ reduces the number of potential candidates to 29.

The procedure was then repeated at unit-cell volumes in the $\pm 7\%$ range of the equilibrium volume. The volume dependence of the descriptor, i.e. $zT(V)$, is evaluated and we identify $zT_{\mathrm{opt}}$ and $zT_0$ which are the magnitude of $zT$ at the volume which maximizes $zT(V)$ and at the calculated equilibrium volume respectively. Thus, for any given material, if $\frac{zT (V_{\mathrm{opt}})}{zT(V_\mathrm{o})} > 1.1$, it will be of interest for the purpose of this work. As can be seen in Fig.~\ref{flowchart} this lowers the number of candidate structures to only eight.

The final step is to consider whether the volume change by alloying is thermodynamically feasible. Consider a mixture of $1-x$ mole fractions of the binary $AB_n$ and $x$ mole fractions of the binary $AC_n$ producing the alloy $AB_{n-x}C_{x}$. The excess energy which is required to obtain the alloy, also referred to as the mixing energy, is
\begin{align}\label{eq:mix}
\Delta E_{\mathrm{mix}} (x;\textrm{AB}_{n},\textrm{AC}_{n}) &= \Delta E_f(AB_{n-x}C_{x})  \nonumber \\
&-\frac{x\Delta E_f(AC_n)+(n-x)\Delta E_f(AB_n)}{n}
\end{align}
where $\Delta E_f$ are the formation energies per atom for the given compounds. We have calculated the mixing energies from Eq.~\eqref{eq:mix}, by taking Special Quasi-random Structure (SQS) alloy distributions for different values of $x$ using the ATAT code \cite{avdw:atat}. For solids, since at ambient pressures, $P\Delta_{\mathrm{mix}}V$ term is negligible, the enthalpy of mixing can be approximated as $\Delta H_{\mathrm{mix}}=\Delta E_{\mathrm{mix}}$. Thus the mixing Gibbs free energy is,
\begin{align}\label{eq:gibbs}
\Delta G_{\mathrm{mix}}(x)&= \Delta E_{\mathrm{mix}}(x)  \nonumber \\
&+\frac{RT}{n+1} \big[ x\ln x+(1-x)\ln (1-x) \big]
\end{align}
where, the last term accounts for entropy of mixing. For each of the alloys, the minimum temperatures, at which the entropy gain in alloy formation compensates the energy cost of mixing, can be obtained by minimizing Eq.~\eqref{eq:gibbs}. We will report $T_m$, which is the maximum temperature of the boundary of corresponding miscibility. As will be discussed later, this in the end limits the number of potential candidates to only four.

\subsection{Identifying Mg$_2$X as a promising candidate.}

\begin{figure}[b]
%%\centerline{\epsfxsize=0.95\textwidth\epsffile{Paper4_Fig4}}
\centering
\includegraphics[width=9cm, clip=true]{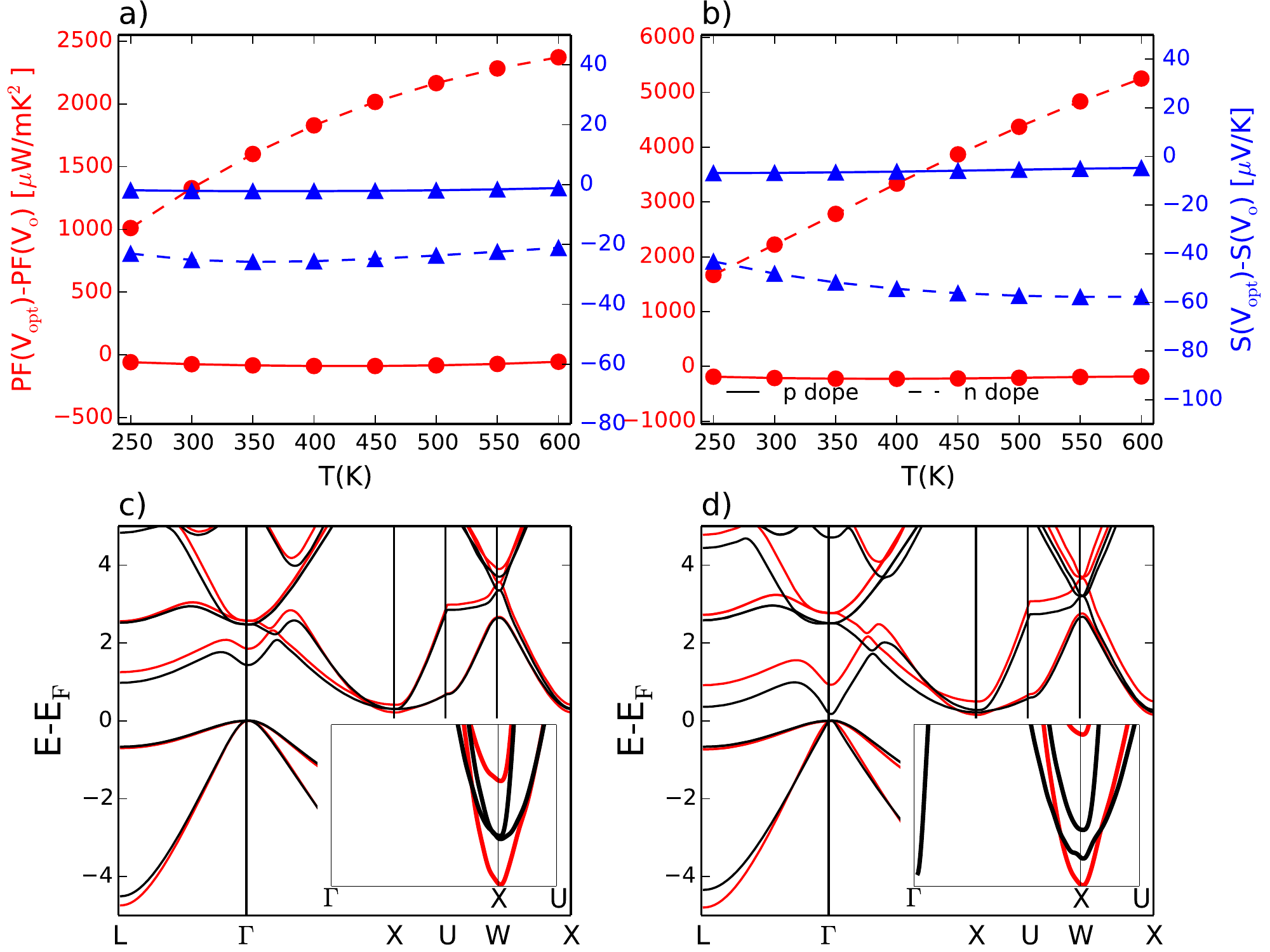}
\caption{Volumetric Band Alignment in Mg$_2$X. In the top panel, Fig.~a and Fig.~b respectively show the change in $PF(V_{\mathrm{opt}})-PF(V_0)$ (in red circles) and $S(V_{\mathrm{opt}})-S(V_0)$ (blue triangles), as a function of temperature at $n=2\times10^{20}$ cm$^{-3}$. Furthermore, in both the graphs the dashed lines correspond to $n-$doping while the bold line represents the data for the $p-$doping scenario. The bottom panels, Fig.~c and Fig.~d respectively show the bandstructures for Mg$_2$Si and Mg$_2$Ge, at $V_0$ (red) and $V_{\mathrm{opt}}$ (black). The insets represent the zoomed-in view of the conduction band minimum. The conventions used in this figure will be used in the forthcoming figures presenting similar results for other candidates in this paper.}
\label{Fig3}
\end{figure}

In the following we will introduce the HT scheme by example, using the Mg$_2$Si/Ge/Sn system, which is well known for its potential VBA.\cite{mg2si_paper}
In the top panel of Fig.~\ref{Fig3}, the temperature dependance of the enhancement of the TE properties upon volume optimization, i.e. $PF(V_{\mathrm{opt}})-PF(V_0)$ and $S(V_{\mathrm{opt}})-S(V_0)$ for the anti-fluorite structure of Mg$_2$Si (Fig.~\ref{Fig3}a) and Mg$_2$Ge (Fig.~\ref{Fig3}b) are shown at a doping of $n=2\times10^{20}$ cm$^{-3}$. In the bottom panel Fig.~\ref{Fig3}c, the bandstructures for Mg$_2$Si at equillibrium volume and the optimized volume are plotted. Likewise, Fig.~\ref{Fig3}d illustrates the bandstructures at $V_0$ and $V_{\mathrm{opt}}$ for Mg$_2$Ge. 

%\begin{figure}
%\includegraphics[width=.30\textwidth]{mg2si_fermi}
%\caption{ The BZ of Mg$_2$Si, illustrating the electron pocket (in blue) at the X point for $V_{\mathrm{opt}}$. There exist no electronic pockets (i.e. a vanishing Fermi surface) for $V_0$.}
%\label{mg2si_fermi}
%\end{figure}

%In accordance to this, carrier concentrations between $\approx 10^{19}-10^{20}$~cm$^{-3}$ can be achieved.\cite{mg2si_defect} Because of the similar chemistry between Si and Ge, the same can be expected for Mg$_2$Ge. In practise, Al can be used as an effective $n-$dopant for Mg$_2$X systems \cite{Mg2Si_Al}.

\begin{figure}
\includegraphics[width=.45\textwidth]{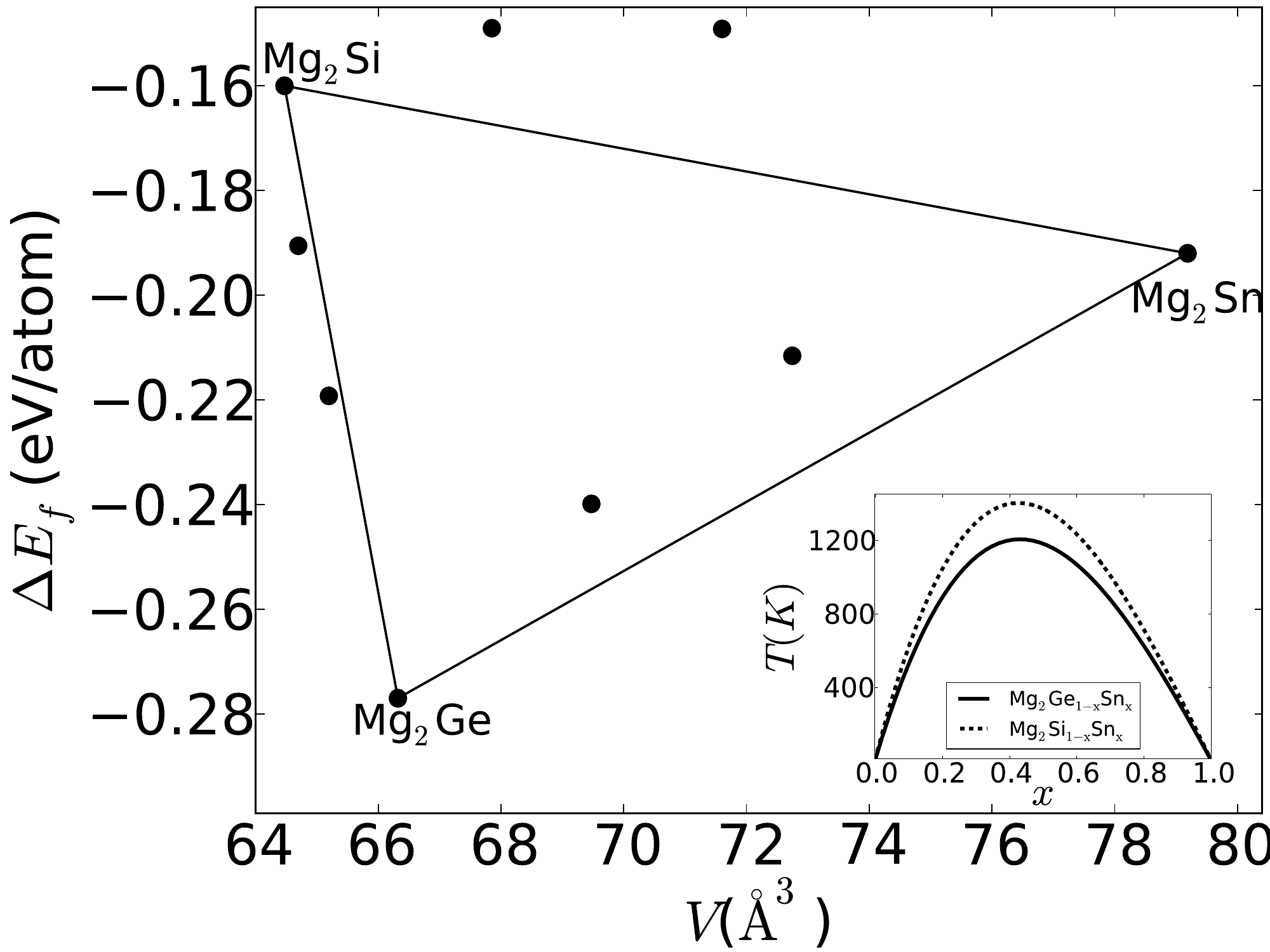}
\caption{Formation energy vs volume for Mg$_2$X$_{1-x}$Y$_{x}$ alloys. Here X,Y are Si, Ge and Sn.  The insert shows the miscibility gap calculated by minimizing Eq.~\eqref{eq:gibbs}. }
\label{triag_Mg2X}
\end{figure} 

In Fig.~\ref{Fig3}a and Fig.~\ref{Fig3}b we observe that for the $p-$type behavior the change in PF with volume is negligible for all $T$. On the other hand, for n-type behavior, the $PF$ shows a pronounced increase at the optimized volume for both the compounds, that further enhances with $T$. Moreover $S$ also exhibits an increase. Note that for $n-$doping, the higher the negative value of $S(V_{\mathrm{opt}})-S(V_0)$, the more is the desired enhancement of the thermopower and $zT$. Also, it is well known that Mg$_2$Si exhibits a persistent $n-$type conductivity under any kind of crystal growth conditions.\cite{mg2si1,mg2si2,mg2si3}. 
When we analyze the variation in the conduction bands (CB) of the compounds with volume (Fig.~\ref{Fig3}c and Fig.~\ref{Fig3}d), the large increment in the PFs for $n-$doped Mg$_2$Si and Mg$_2$Ge becomes quite apparent. At the X point in the BZ near the conduction band minimum (CBM), the first and second CBs directly coincide at $V_{\mathrm{opt}}$ for Mg$_2$Si and are only a few meV appart for Mg$_2$Ge (see insets). Therefore, even at low dopings, the charge carriers (electrons) residing in both the bands will contribute to enhance the PF and Seebeck coefficient, as opposed to the contributions coming from electrons in only a single band, at equillibrium volume. 

For Mg$_2$Si, at a doping of $n=2\times 10^{20}$ cm$^{-3}$ and $T=600$~K, the highest increment in $zT$ is observed at volume increase of 2\% from the equilibrium volume, Table~\ref{Table1}. According to Vegard's law this corresponds roughly to alloying Mg$_2$Si with $10\%$ Mg$_2$Sn. This value will also be dependent on the carrier concentration and temperature at which the transport properties are calculated. The important rule is if a large VBA effect is observed at a small volume change that in principle can be attained by alloying. The second crucial criterion is whether the alloying is thermodynamically feasible. The energy of formation vs volume for Mg$_2$X$_{1-x}$Y$_{x}$ solid solution is depicted in Fig.~\ref{triag_Mg2X}. Here X is Si or Ge while Y is Ge or Sn. It can be seen that the excess mixing energy required for the formation of stable alloys is small in magnitude. As a result the optimal Sn content can be reached at moderate growth temperature, Fig.~\ref{triag_Mg2X}insert. $\Delta E_{\mathrm{mix}}$ and $T_m$ are in good agreement with previous DFT calculation\cite{Viennois2012}, especially considering that we employ a somewhat different computational approach using SQS based supercells and linear interpolation between the calculated points when minimizing Eq.~\eqref{eq:gibbs}. We also find a good agreement with CALPHAD and experimental results for this system.\cite{Jung2007192,Jung2010,Kozlov20113326}.

Thus in the particular cases of antifluorite Mg$_2$Si and Mg$_2$Ge, the VBA can be conveniently achieved through alloying. While Mg$_2$Si$_{1-x}$Ge$_{x}$ alloy thermodynamically exists as a solid solution even at low temperatures, for Mg$_2$Si$_{1-x}$Sn$_{x}$ and Mg$_2$Ge$_{1-x}$Sn$_{x}$ alloys the desired volume increment can be achieved at the expense of only a small magnitude of mixing energy \cite{Viennois2012,Jung2007192,Jung2010,Kozlov20113326}.

\section{Results and discussion}

\begin{figure}[!ht]
{\includegraphics[width=0.51\textwidth]{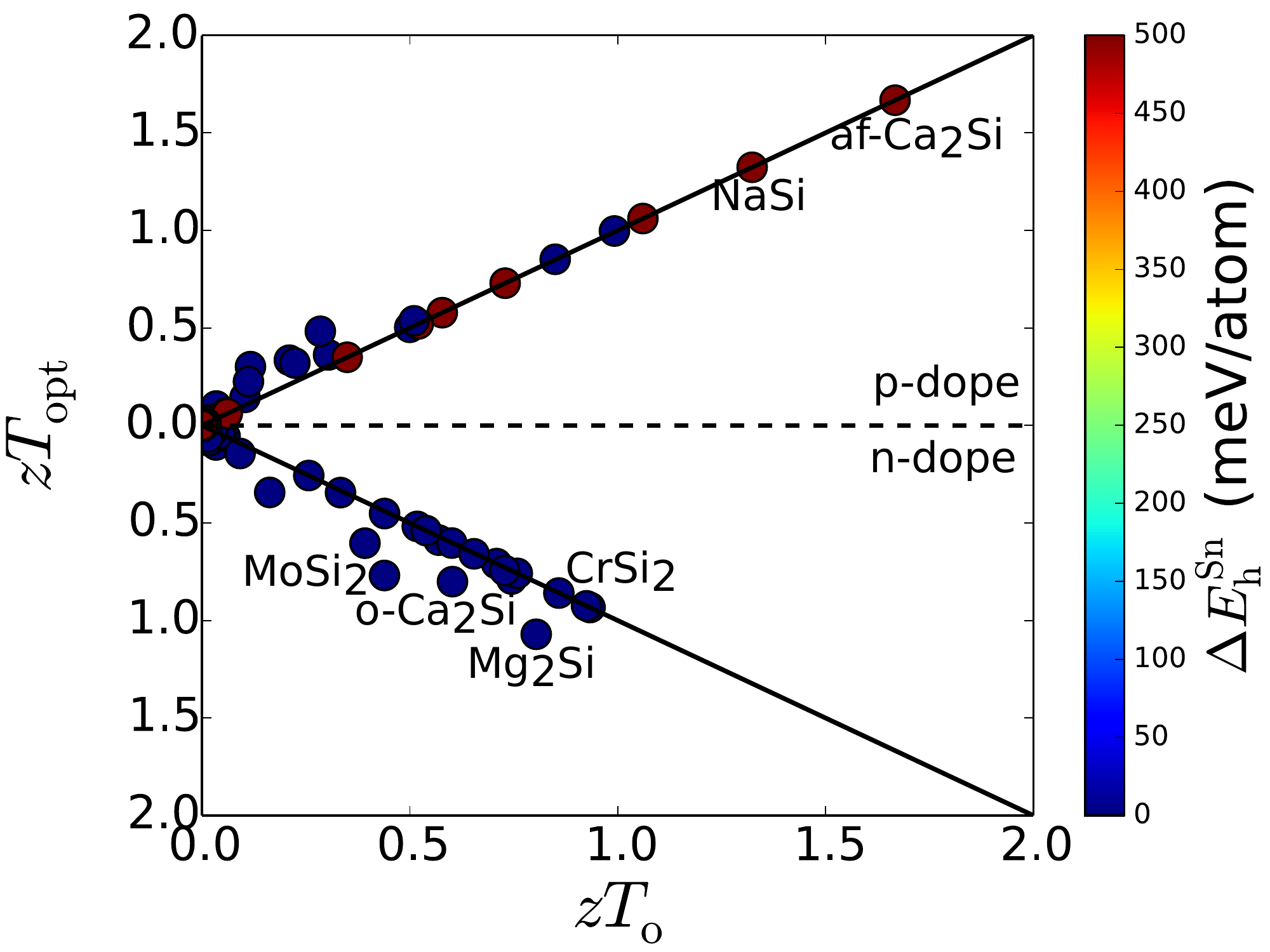}}
\hfill
{\includegraphics[width=0.51\textwidth]{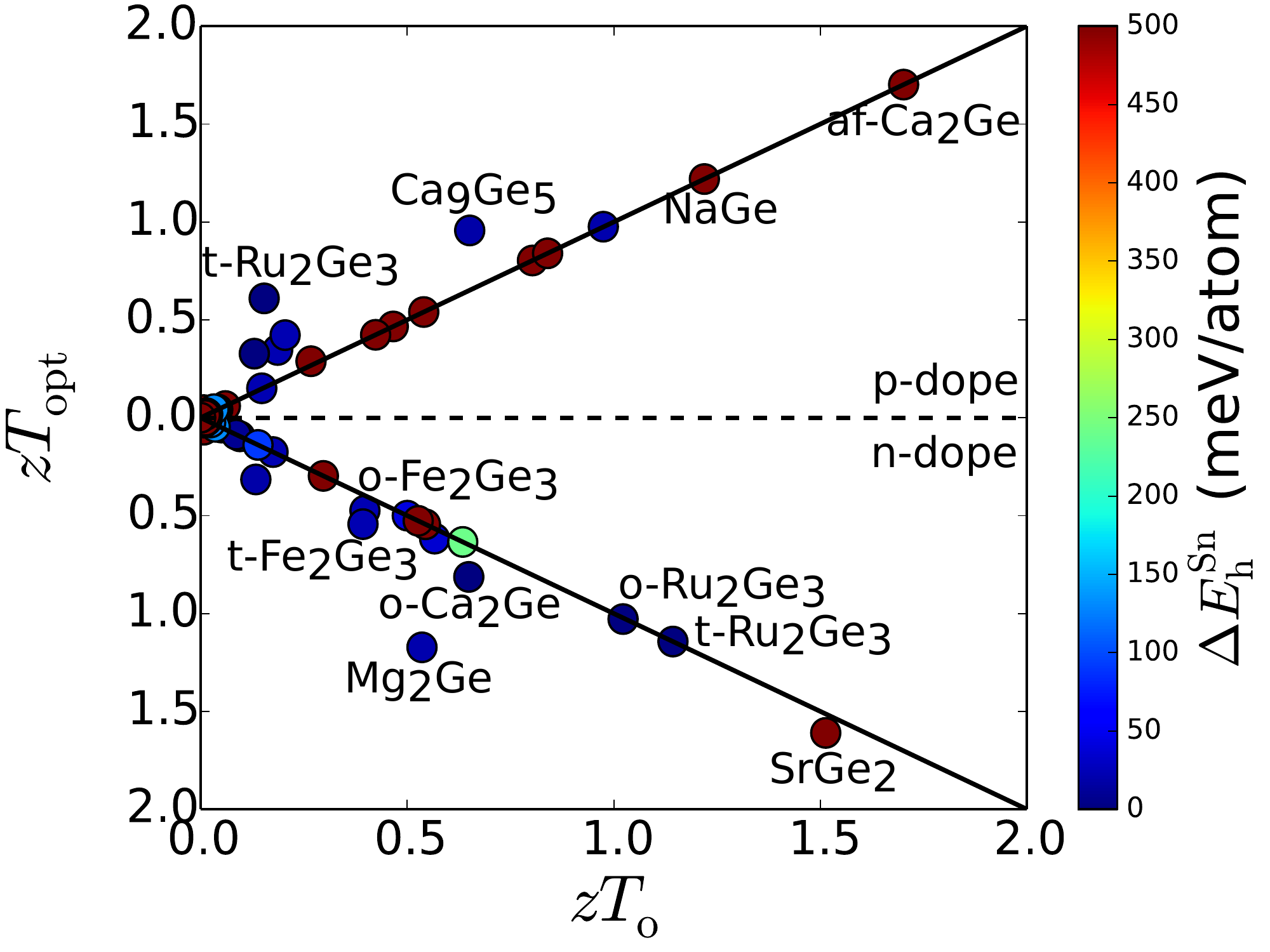}}
\caption{$zT_\mathrm{opt}$ vs $zT_\mathrm{o}$ for all silicides (\textit{top}) and germanides (\textit{bottom}) at doping $n=2\times10^{20}$~cm$^{-3}$ and $T=600$ K. Furthermore, the scatter points are colored based on the distance from the convex hull, $\Delta E_{\mathrm{h}}^{\mathrm{Sn}}$, of the corresponding Sn compound with the same structure as the host binary. For example, in the case of the Ca$_{9}$Ge$_{5}$ compound, $\Delta E_{\mathrm{h}}$ of the corresponding iso-structural Ca$_{9}$Sn$_{5}$ is used for the color code, and so on. The candidates that exhibit VBA, $\frac{zT_\mathrm{opt}}{zT_0}>1.1$, or have a large value of $zT_0$ are labelled. Also, if a compound crystallizes in more than one phase, the corresponding structure is also indicated with a prefix \textit{o-} for orthorhombic, \textit{t-} for tetragonal and \textit{af-} for anti-fluorite phase.}
\label{fig:ZT}
\end{figure}

\subsection{Electronic screening}
Based on the descriptors discussed above other candidates for VBA effect will now be identified.
The top and bottom panels in Fig.~\ref{fig:ZT} illustrates $zT_\mathrm{opt}$ vs $zT_0$ respectively for all silicides and germanides, at a doping of $n=2\times10^{20}$~cm$^{-3}$ and $T=600$~K. Each data point in Fig.~\ref{fig:ZT} represents a particular compound. %Note that in Fig.~\ref{fig:ZT}, if any of the appropriate alloying candidates do not exist, the distance from the convex hull is set at a very large value of 500 meV. In other words, the formation of that particular alloy would be energetically impossible. 

%%450K%%

%\begin{table}
%\begin{tabular}{ |c|c|c|c|c|c| }
% \hline
% Compound & $zT_{\mathrm{opt}}$ & $zT_{\mathrm{opt}}/zT_0$ & $\Delta E_{\mathrm{hull}}$ (meV) & doping & $V_{opt}$ (\%)\\
% \hline
%  af Mg$_2$Si&0.69 &1.50 &0 &$n$& \\
%  af Mg$_2$Ge&0.75 &2.57 &0 &$n$& \\
%  orth Ca$_2$Si& 0.47 &1.35 &0 &$n$& \\
%  orth Ca$_2$Ge&0.48 &1.28 &0 &$n$& \\
%  hex Ca$_{9}$Ge$_{5}$&1.11 & 1.60 &20.0&$p$& \\
%  hex $\beta-$MoSi$_2$& 1.07 & 3.84 & 0.0 & $n$& \\
%  orth Fe$_2$Ge$_3$& 0.55 & 1.10 & 22.0 & $n$& \\
%  tet Fe$_2$Ge$_3$& 0.58 & 1.18 & 20.0 & $n$& \\
%  tet Ru$_2$Ge$_3$& 0.32 & 4.69 & 0 & $p$& \\
% \hline
%\end{tabular}
%\caption{$T=450$~K.}
%\label{Table1}
%\end{table}

\begin{table}
\begin{tabular}{ |c|c|c|c|c|c| }
 \hline
 Compound & $zT_{\mathrm{opt}}$ & $zT_{\mathrm{opt}}/zT_0$ & $\Delta E_{\mathrm{h}}$ [$\Delta E_{\mathrm{h}}^{\mathrm{Sn}}$] & doping & $V_{opt}$\\
  &  &  & (meV/atom) &  & (\%)\\
 \hline
  Mg$_2$Si&1.07 &1.33 &0[0] &$n$&2.0 \\
  Mg$_2$Ge&1.17 &2.19 &0[0] &$n$&5.0 \\
  Ca$_2$Si& 0.80 &1.33 &0[0] &$n$&5.0 \\
  Ca$_2$Ge&0.81 &1.25 &0[0] &$n$&2.5 \\
  Ca$_{9}$Ge$_{5}$&0.96 & 1.47 &37.4[17.9]&$p$&6.1 \\
  $\beta-$MoSi$_2$& 0.77 & 1.75 & 27.3[180.8] & $n$&3.0 \\
  o-Fe$_2$Ge$_3$& 0.47 & 1.19 & 0.1[22.5] & $n$&3.0 \\
  t-Fe$_2$Ge$_3$& 0.54 & 1.38 & 0[22.8] & $n$&3.0 \\
 \hline
\end{tabular}
\caption{Summary of eight candidates that exhibit the VBA effect. We summarize $zT_{\mathrm{opt}}$ and $zT_{\mathrm{opt}}/zT_0$ magnitudes at doping of $n=2\times10^{20}$ cm$^{-3}$ and $T=600$~K. The distance from the convex hulls, $\Delta E_{\mathrm{h}}$ [$\Delta E_{\mathrm{h}}^{\mathrm{Sn}}$] of the candidate [alloying possibility], carrier sign and $V_{opt}$ (expressed as an increment from $V_0$) are also summarized. }
\label{Table1}
\end{table}

Thus Fig.~\ref{fig:ZT} contains three vital descriptors to identify candidates for VBA, (i), the magnitude of $zT_0$, which indicates if a particular candidate is a good TE material, (ii), the value $\frac{zT_\mathrm{opt}}{zT_0}$, which helps us to identify the promising candidates exhibiting the volumetric enhancement of their TE properties, (iii) the magnitude of the distance from the corresponding convex hull of the possible alloying choices, which would produce the necessary volumetric change upon alloying (according to Vegard's law). 

A summary of the promising silicide and germanide structures that exhibit a large VBA effect are shown in Table~\ref{Table1}, with the corresponding value of $zT_{\mathrm{opt}}$, $zT_{\mathrm{opt}}/zT_0$ and $\Delta E_{\mathrm{h}}$ [$\Delta E_{\mathrm{h}}^{\mathrm{Sn}}$]. Eight such compounds were identified to exhibit encouraging VBA effects upon using the magnitude of $zT_{\mathrm{opt}}/zT_0$ and $\Delta E_{\mathrm{h}}$ as descriptors.

In general, with increasing doping we observe that more number of candidates show the VBA effect. This is due to a large value of $zT$ itself that is improved upon doping. However the best candidates, as listed in Table~\ref{Table1}, exhibiting VBA remained unchanged. Apart from Mg$_2$Si and Mg$_2$Ge, discussed previously, orthorhombic Ca$_2$Si and Ca$_2$Ge, hexagonal Ca$_9$Ge$_5$ and hexagonal-MoSi$_2$ show encouraging results. 

Note that in Fig.~\ref{fig:ZT} there are compounds that show very little to no VBA effect but exhibit large $zT_0$. Indeed, despite our semi-emperical determination of $zT$, we correctly predict most of these binaries with large values of $zT_0$, that are already established as encouraging TE materials.\cite{fedorov} Compounds which exhibit $zT_0$ in excess of 1.0, at  $n=2\times10^{20}$ carriers per cm$^{-3}$, are the orthorhombic and tetragonal structures of Ru$_2$Ge$_3$, cubic RuSi (at low temperatures) and tetragonal NaGe and NaSi. Furthermore, the following compounds have $zT_0$ between 0.5 and 1.0 (not listed in Fig.~\ref{fig:ZT}): hexagonal CrSi$_2$, Mn$_4$Si$_7$, FeSi$_2$, ReSi$_{1.75}$, and CoSi. 

\subsection{Feasibility in alloying}

\begin{table}
\begin{tabular}{ |c|c|c|c|c| }
 \hline
 Alloys & $\mathrm{x}_{\mathrm{alloy}}$ & $\Delta E_{\mathrm{mix}}(x=0.25)$  & $T_m$ (K) & $\mathrm{x}$ \\
 & &(kJ/mol.atom) & & $(T=800~\mathrm{K})$ \\
 
 \hline
  Mg$_2$Si$_{\mathrm{1-x}}$Sn$_{\mathrm{x}}$&0.09 &1.997 &1281&0.171 \\
  Mg$_2$Ge$_{\mathrm{1-x}}$Sn$_{\mathrm{x}}$&0.26 &1.595 &1023&0.138  \\
 \hline
  Ca$_2$Si$_{\mathrm{1-x}}$Sn$_{\mathrm{x}}$&0.30 &0.013 &78& all \\
  Ca$_2$Ge$_{\mathrm{1-x}}$Sn$_{\mathrm{x}}$&0.18 &0.015 & 97& all \\
  \hline
  Ca$_9$Ge$_{\mathrm{5-x}}$Sn$_{\mathrm{x}}$&0.28 & 4.495& $>5000$&  0.019 \\
  \hline
  $\beta-$MoSi$_{\mathrm{2-x}}$Sn$_{\mathrm{x}}$&0.07 & 30.980& $>5000$& 0.008 \\
  \hline
  o-Fe$_2$Ge$_{\mathrm{3-x}}$Sn$_{\mathrm{x}}$&0.10 &32.746 &$>5000$& 0.004 \\
  \hline
  t-Fe$_2$Ge$_{\mathrm{3-x}}$Sn$_{\mathrm{x}}$&0.10 &29.681 &$>5000$& 0.005 \\
 \hline
\end{tabular}
\caption{\noindent Alloy thermodynamics of the candidates. $\mathrm{x}_{\mathrm{alloy}}$ is the amount of alloying component, calculated using Vegard's law, required to obtain the volume change for the VBA effect. $\Delta E_{\mathrm{mix}}(x=0.25)$ is the mixing energy required to form the alloy (at $\mathrm{x}=0.25$). $T_m$ is the maximum temperature of the boundary of the miscibility gap and $x(T=800~\mathrm{K})$ is the amount of alloying component which can be accomodated at $T=800$~K in the alloy.}
\label{table2}
\end{table}
We will now explore the posibility of forming thermodynamically stable alloys among the candidates for VBA, identified based on the descriptors discussed so far. These alloys are summarized in Table.~\ref{table2} together with proportion of alloying component $\mathrm{x}_{\mathrm{alloy}}$ required to produce the highest optimized PFs in the candidates. Note that the optimal VBA for most candidates are achieved at $x\leq0.30$. %The alloying possibilities for each candidates considered in Table.~\ref{table2} correspond to the simplest possible choice. For example, the volume of MSi compound could potentially be increased with MSi$_{1-x}$Sn$_{x}$ alloys. The reason for this choice is because MSn is likely to have a larger volume in comparision to its isostructural MSi.
%The next three columns of Table.~\ref{table2} contain the mixing energy $\Delta E_{\mathrm{mix}}$ for $x=0.25$, temperature which closes the miscibility gap of the alloying components, i.e. $T_m$ and the amount of foreign alloy that can be accommodated at $T=800$~K. 

\begin{figure}
\includegraphics[width=.5\textwidth]{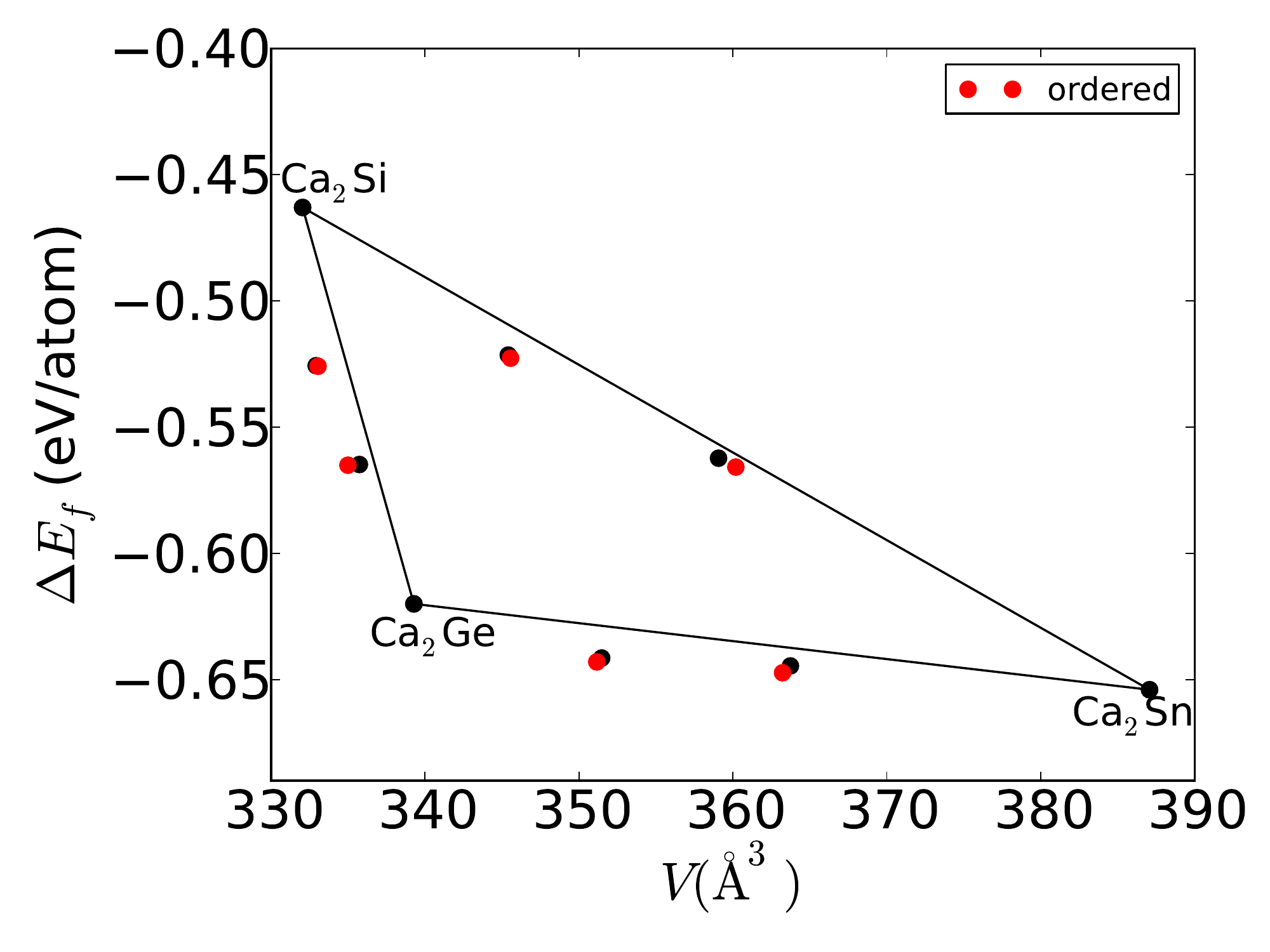}
\caption{Formation energy vs volume for Ca$_2$X$_{1-x}$Y$_{x}$ solid solutions (X and Y are Si, Ge and Sn).}
\label{triag_Ca2X}
\end{figure}

\begin{figure}
\includegraphics[width=.4\textwidth]{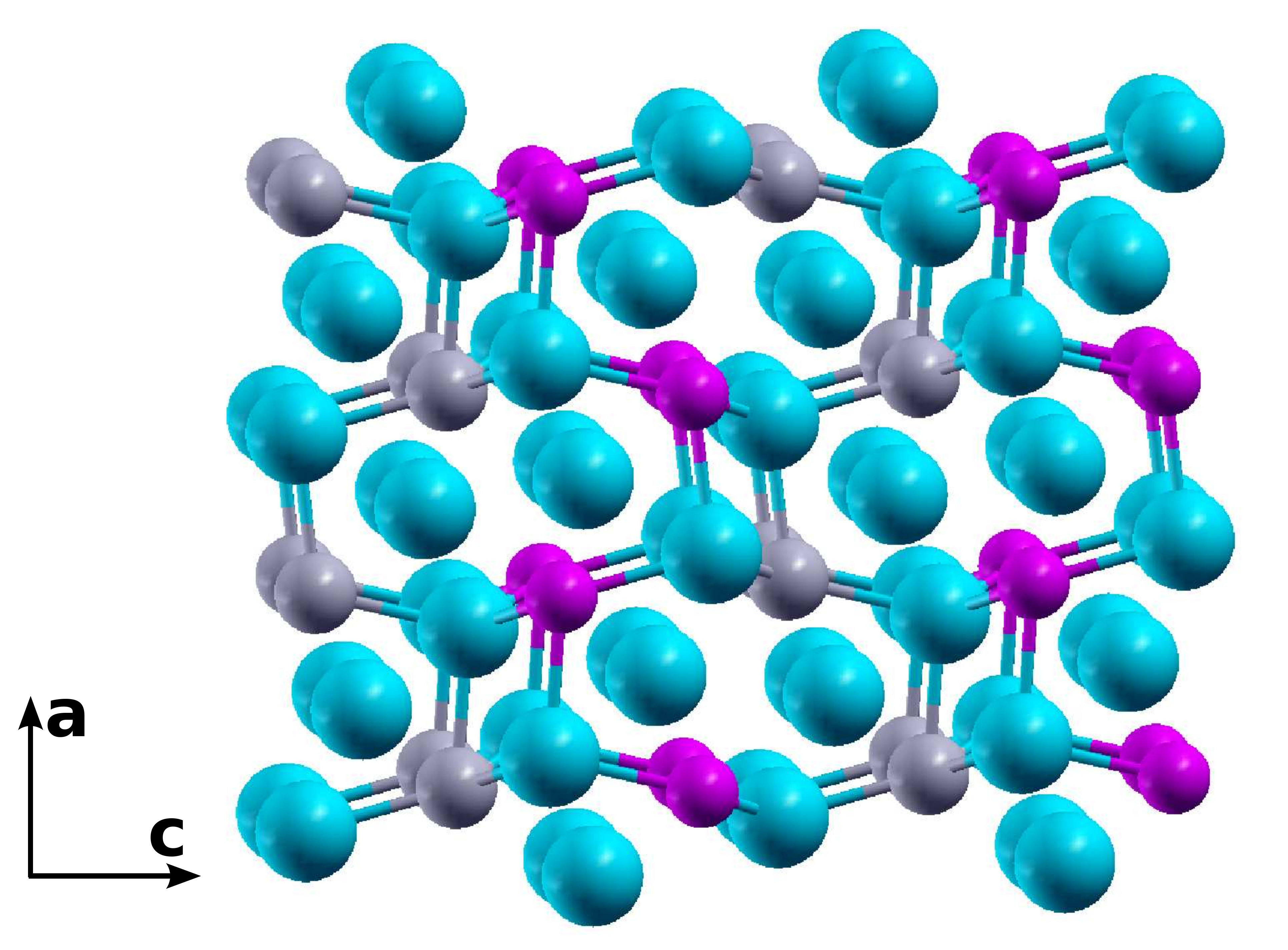}
\caption{Ordered alloy structure is shown here for a $2\times2\times2$ supercell. Here the blue balls are Ca atoms, the grey balls are X atoms and the magenta ones are Y atoms. Note that an unitcell contains 12 atoms, i.e. Ca$_8$X$_2$Y$_2$ composition. The shortest crystallographic axis, \textbf{b}, is perpendicular to the plane. This structure is related to the Co$_2$Si (or TiNiSi) structure type with space group $Pnma$ \cite{structure}.}
\label{ordered}
\end{figure} 

In the previous section we have established that Ca$_2$Si and Ca$_2$Ge are attractive candidates for VBA effect based on the descriptors which characterize their transport properties. In Fig.~\ref{triag_Ca2X}, we illustrate the energy of formation of orthorhombic Ca$_2$X$_{1-x}$Y$_{x}$ solid solutions as a function of volume. We observe that for $x=0.25$ and $x=0.5$, the SQS generated random structures are actually lower in energy in comparision to the parent structures. We therefore generated all the possible ordered structures of Ca$_{64}$Si$_{24}$Sn$_{8}$ and Ca$_{64}$Si$_{16}$Sn$_{16}$ within a $2\times 2\times 2$ supercell. We found a previously unreported ordered structure in which the Sn atoms order along the short ${\bf b}$-axis to be marginally lower in energy in comparision to the random SQS structure chosen. Fig.~\ref{ordered} illustrates this ordered structure, which can be thought of as being stabilized by Sn-Sn bonding along the short axis of the orthorhombic cell and by the shortest Sn-Sn distance within the unitcell.  In Table.~\ref{table2} the energy difference between the ordered and disordered alloy is reported as $\Delta E_{\mathrm{mix}}$. The energy difference is very low and the alloying necessary to obtain the desired volume increment should be readily achieved through Ca$_2$Si$_{\mathrm{1-x}}$Sn$_{\mathrm{x}}$ and Ca$_2$Ge$_{\mathrm{1-x}}$Sn$_{\mathrm{x}}$ alloys. Thus Ca$_2$Si and Ca$_2$Ge compounds are confirmed to exhibit the VBA effect and the required volume change can also be conviniently achieved through alloying. Interestingly, Ca$_2$Sn has been previously proposed \cite{Parkerorth} as a thermoelectric material.

Unfortunately for all the other alloys, we obtained the mixing energy $\Delta E_{\mathrm{mix}}$ to be very large (see Table.~\ref{table2}). Therefore for Ca$_9$Ge$_5$, $\beta$-MoSi$_2$, o-Fe$_2$Ge$_3$ and t-Fe$_2$Ge$_3$, the optimal VBA volume cannot be expected to be attained by the simple alloy choices listed in Table.~\ref{table2}. 

%We had previously discussed the thermodynamic stability of the corresponding candidate alloys based on the distance from the convex hull reported in Fig.~\ref{fig:ZT}.

Interestingly, we observe that for all four alloys having a large $\Delta E_{\mathrm{mix}}(x=0.25)$ in Table~\ref{table2}, the corresponding pure compounds are found above the convex hull, Table~\ref{Table1}. E.g. in the case of $\beta$-MoSi$_2$ we had obtained that $\beta$-MoSn$_2$ is far from the convex hull (i.e. $\approx180$ meV/atom). 
 This could indicate that $\Delta E_h^{\textrm{Sn}}$ in Fig.~\ref{Fig3}, can serve as a minimal requirement for the possibility of forming a stable alloy system. Such a descriptor could help to avoid the computationally more expensive SQS supercell calculations.
 
\subsection{Origin of high $zT (V_{\mathrm{opt}})$ in encouraging structures}
In the following, we shall investigate the role played by the electronic structure on VBA effect for the candidates. The main focus will be put on CaSi$_2$ and CaGe$_2$ where it has been shown that the necesarry volume expansion should be achieved by alloying. We will furthermore discuss $\beta$-MoSi$_2$ and Ca$_9$Ge$_5$. For these candidates, while the optimal VBA volume can not be attained by alloying, we point out that even a small volume increase generated from either alloying or thermal expansion could improve their transport properties through the VBA effect.

\subsubsection{Orthorhombic Ca$_2$Si and Ca$_2$Ge}

\begin{figure}[ht]
%%\centerline{\epsfxsize=0.95\textwidth\epsffile{Paper4_Fig4}}
\centering
\includegraphics[width=9cm, clip=true]{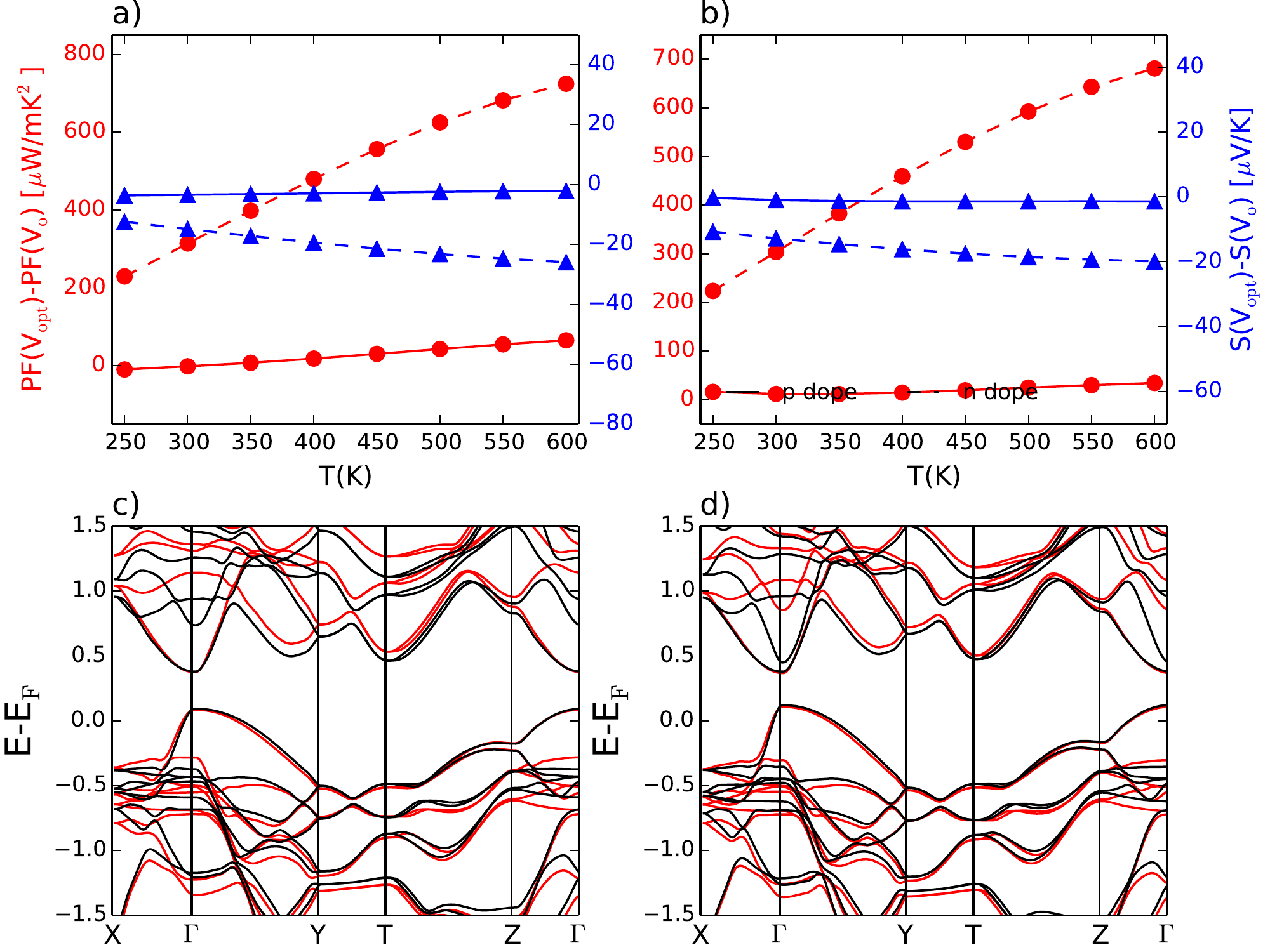}
\caption{The temperature dependance of $PF(V_{\mathrm{opt}})-PF(V_0)$ and $S(V_{\mathrm{opt}})-S(V_0)$ for Ca$_2$Si and Ca$_2$Ge respectively in  Fig.~a and Fig~b. Furthermore, in Fig.~c and Fig.~d we depict the corresponding bandstructures ($V_{\mathrm{opt}}$ in black and $V_{\mathrm{o}}$ in red; Fig.~c for Ca$_2$Si and Fig.~d for Ca$_2$Ge). The orthorhombic structure for these compounds are thermodynamically stable. Both the $n-$doped compounds exhibit an increase in $PF$  and $S$ that can be justified by the increased DOS around the CB manifold. Note also the close band alignment of the first two CBs at the $\Gamma$ point in BZ, for Ca$_2$Ge. }\label{Fig5}
\end{figure}

%Ca$_2$X compounds crystallize in an orthorhombic structure, located on their respective binary convex hulls. 
Both Ca$_2$X compounds show a large volumetric enhancement in their TE properties. The top panel in Fig.~\ref{Fig5} illustrate the temperature dependance of the difference in $PF$ (and $S$) between $V_{\mathrm{opt}}$ and $V_0$, for Ca$_2$Si in Fig.~\ref{Fig5}a and Ca$_2$Ge in Fig.~\ref{Fig5}b. Likewise, the bottom panel shows the bandstructures of the two compounds (Fig.~\ref{Fig5}c for Ca$_2$Si and Fig.~\ref{Fig5}d for Ca$_2$Ge), at the two volumes.

\begin{figure}
\includegraphics[width=.35\textwidth]{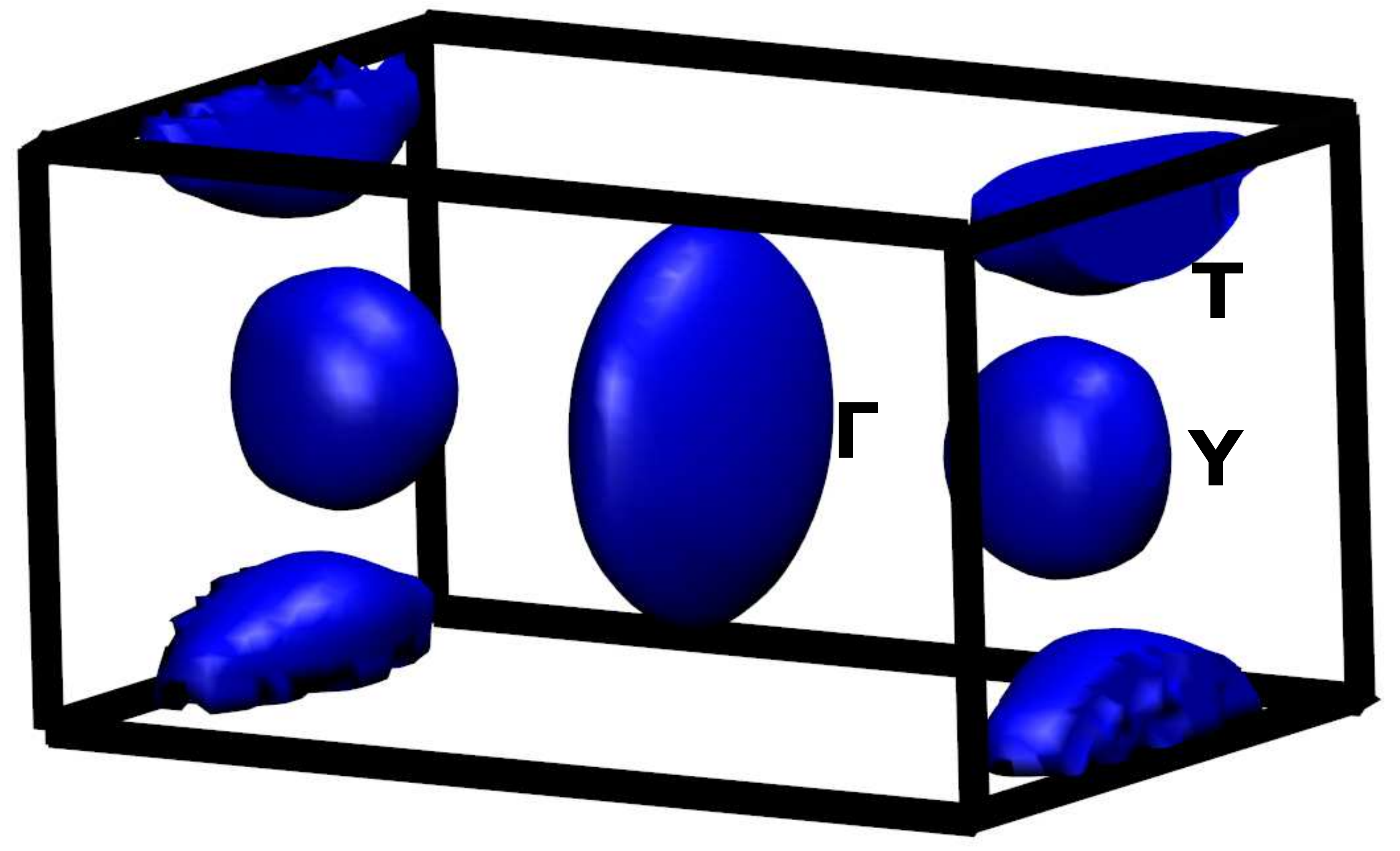}
\caption{ The BZ of orthorhombic Ca$_2$Si at $V_{\mathrm{opt}}$, illustrating the electron pockets (in blue) at the high symmetry T and $\Gamma$ points as well as the low symmetry pocket around Y. The electron pocket around the Y point does not exist at $V_0$.}
\label{ca2si_fermi}
\end{figure}

The $p-$type behavior for both the compounds show a negligible amount of increase in $PF$ and $S$ at $V_{\mathrm{opt}}$. Accordingly, the VBM, around the vicinity of $\Gamma$ point, is relatively unchanged. 
The situation is different for the CBM which has contributions coming from the $\Gamma$ and T points at $V_0$. Due to this we have $zT^{\mathrm{Ca_2Si}}_0=0.60$ and $zT^{\mathrm{Ca_2Ge}}_0=0.64$, at $T=600$ K and $n=2 \times 10^{20}$ cm$^{-3}$. At optimal volume, n-type Ca$_2$Si and Ca$_2$Ge show an increase in their TE properties to $zT^{\mathrm{Ca_2Si}}_{\mathrm{opt}}=0.80$ and $zT^{\mathrm{Ca_2Ge}}_{\mathrm{opt}}=0.81$. 
From the bandstructures of Ca$_2$Si in Fig.~\ref{Fig5}c, one can understand this volumetric enhancement. There is a lowering of the CBM at the T and around the Y point. This aligns several pockets close to the CBM edge, which is further illustrated in Fig.~\ref{ca2si_fermi}. The constant energy surface of the lowest conduction band of Ca$_2$Si at $V_{\mathrm{opt}}$ shows the electron pockets at the $T$ and $\Gamma$ points as well as those along the $\Gamma-Y$ lines. The latter electron pocket does not exist at $V_0$. In the case of Ca$_2$Ge, a similar lowering of the lowest conduction band along the $\Gamma-Y$ lines is observed in Fig.~\ref{Fig5}d at $V_{\mathrm{opt}}$. Similar carrier pocket shapes have also been identified as the source of high $\sigma$ in isostructural orthorhombic Ca$_2$Pb and Sr$_2$Pb.\cite{Parkerorth}

\subsubsection{Hexagonal Ca$_{9}$Ge$_{5}$}

\begin{figure}[ht]
%%\centerline{\epsfxsize=0.95\textwidth\epsffile{Paper4_Fig4}}
\centering
\includegraphics[width=.5\textwidth]{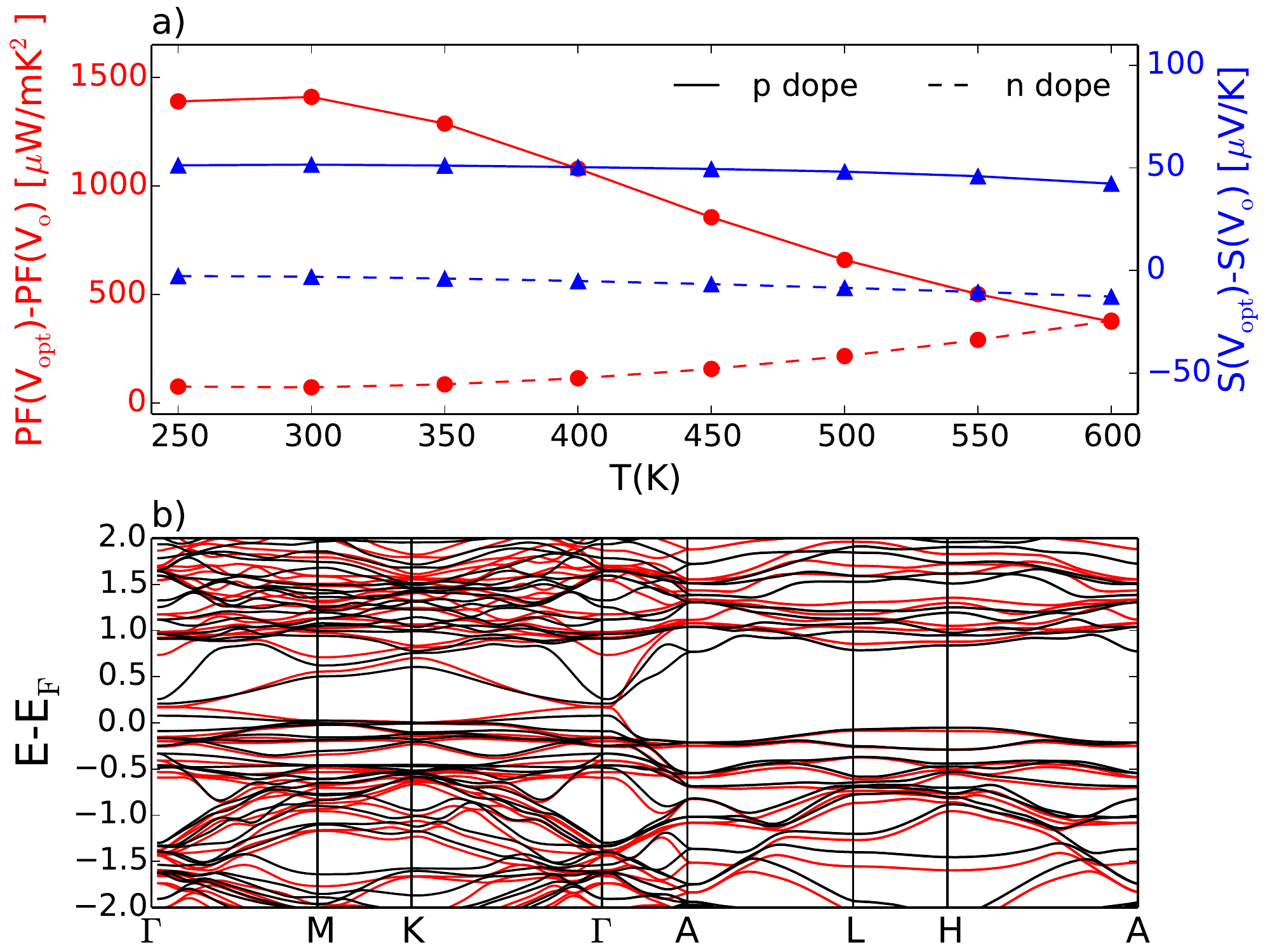}
\caption{$T$ dependence of $PF(V_{\mathrm{opt}})-PF(V_0)$ and $S(V_{\mathrm{opt}})-S(V_0)$ for the hexagonal Ca$_{9}$Ge$_{5}$ (\textit{top}). The corresponding bandstructures at the two volumes are shown in bottom panel. For Ca$_{9}$Ge$_{5}$, there is an opening of the bandgap with volume change that is responsible for the increase in its $PF$ and $S$.}\label{Fig6}
\end{figure}

We shall now discuss the hexagonal structure of Ca$_{9}$Ge$_{5}$ which also exhibits an increase in its TE properties with volume change. In the same fashion as the data presented in the previous sections, Fig.~\ref{Fig6}a shows the temperature depedance of the change in $PF$ and $S$ at the two volumes, $V_{\mathrm{opt}}$ and $V_0$. The bottom panel shows the bandstructure at the two volumes. 

The situation for Ca$_{9}$Ge$_{5}$ is quite interesting. Upon an increase in volume, there is an opening of the bandgap at the $\Gamma$ point in its BZ. This metal to semiconductor transition with volume change is the sole reason behind the enhancement of $PF$ and $S$ at $V_{\mathrm{opt}}$, graphed in Fig.~\ref{Fig6}b. Interestingly, a very high $zT^{\mathrm{Ca_{9}Ge_{5}}}_{\mathrm{opt}}=0.96$ and $zT_{\mathrm{opt}}/zT_0=1.47$ (at $T=600$~K and $n=2 \times 10^{20}$ cm$^{-3}$) was observed.%{\em Sandip: is Ca9Sn5 a metal}

\subsubsection{Hexagonal MoSi$_2$}

\begin{figure}[ht]
%%\centerline{\epsfxsize=0.95\textwidth\epsffile{Paper4_Fig4}}
\centering
\includegraphics[width=.5\textwidth]{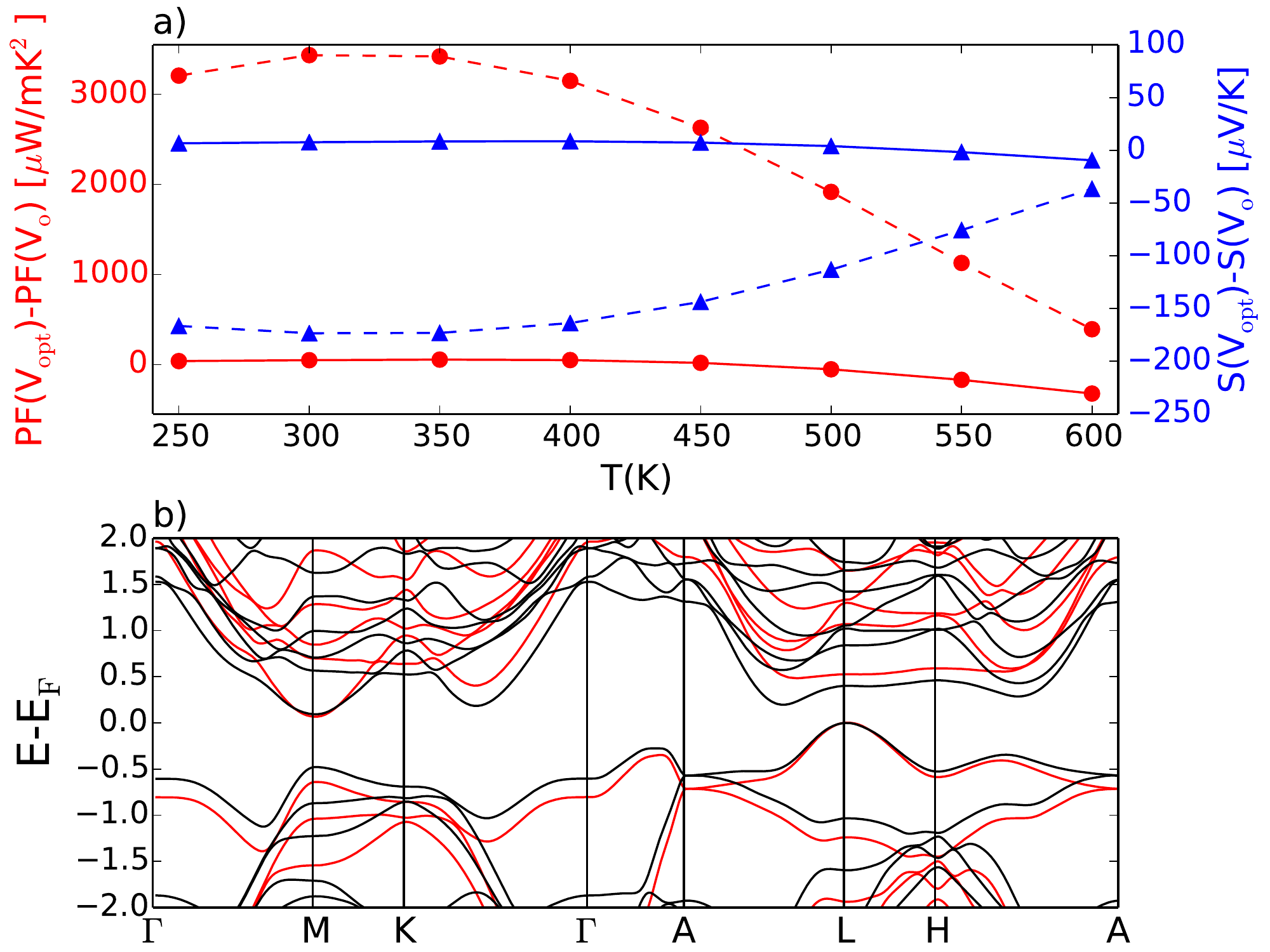}
\caption{We depict the temperature dependence of $PF(V_{\mathrm{opt}})-PF(V_0)$ and $S(V_{\mathrm{opt}})-S(V_0)$ for the $\beta-$MoSi$_2$ in Fig.~a. In the bottom panel (Fig.~b), we present the bandstructures at the two volumes ($V_{\mathrm{opt}}$ in black and $V_{\mathrm{o}}$ in red) for $\beta-$MoSi$_2$. $n-$type MoSi$_2$ shows an appreciable increase in both $PF$ and $S$ at $V_{\mathrm{opt}}$ upto $T=350$ K. This increase can be explained by the lowering and aligning of the first CB manifold along multiple directions in its hexagonal BZ.}
\label{Fig7}
\end{figure}

Molybdenum di-silicide, MoSi$_2$, exists as $\alpha-$MoSi$_2$, having a tetragonal body-centered packing with a space group of $I4/mmm$ and as $\beta-$MoSi$_2$ that has a hexagonal closed packing arrangement with space group $P6_222$. Both the structures are composed of Mo and Si$_2$ layers in which Mo atoms are surrounded by 6 Si atoms. Both the allotropes of MoSi$_2$ exhibit properties such as high melting points, low resistivity and high mechanical strength. 
%$\alpha-$MoSi$_2$ finds extensive applications as refractory ceramic in heating elements, in the electronics industry as gate electrode material and as contact material in microelectronics. 
While $\beta-$MoSi$_2$ is found to be 27.3 meV/atom above the convex hull, it can be conviniently synthesized from Mo and Si powders using spark plasma sintering techniques\cite{mosi2_dft,exp2_mosi2}, thus confirming the $\Delta E_h < 50$~meV/atom criteria. Moreover, since $\alpha-$MoSi$_2$, which is extensively used in microelectronics, can be both $p$ and $n$ doped effectively, the same could be true for the less studied $\beta$ phase.

\begin{figure}
\includegraphics[width=.35\textwidth]{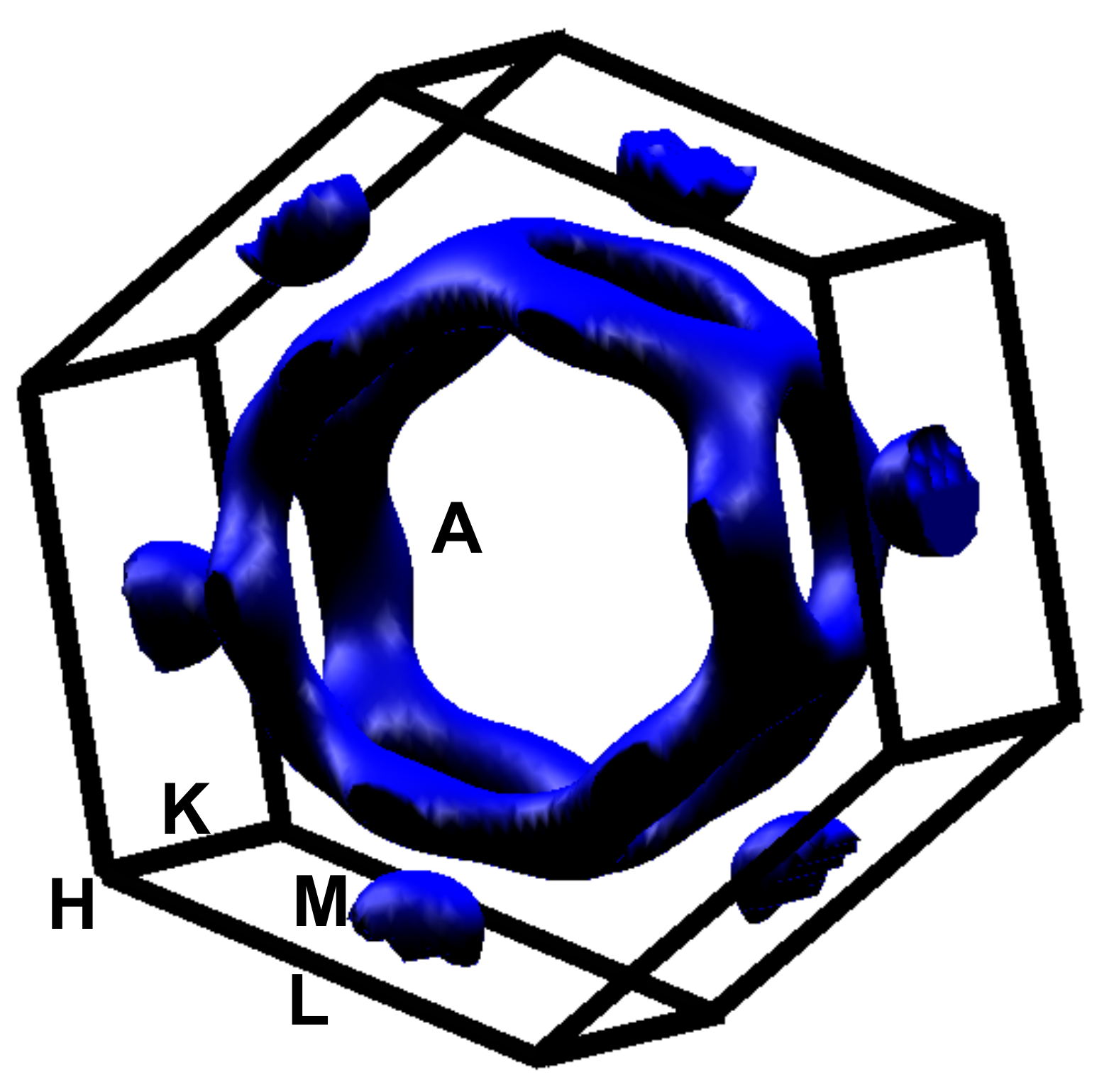}
\caption{ The BZ of $\beta$-MoSi$_2$ illustrating the Fermi surface at $V_{\mathrm{opt}}$. We observe electronic contributions connecting several low symmetry points accross the BZ in the Fermi furface which is responsible for the VBA effect (also see Fig.~\ref{Fig7}b).}
\label{MoSi2_fermi}
\end{figure}

Fig.~\ref{Fig7}a shows the $T$ dependance of difference in the thermoelectric properties ($PF$ and $S$), between $V_{\mathrm{opt}}$ and $V_0$ for $\beta-$MoSi$_2$. Additionally, Fig.~\ref{Fig7}b shows the bandstructures at the two volumes. For n-doped MoSi$_2$ both the $PF$ and $S$ increases drastically by a volume expansion up to $T=350$ K. The enhancement of $PF$ and $S$ is steadily reduced at temperatures above 350~K due to the small band gap. Please note that the results for all the candidates in Table.~\ref{Table1} are presented at $T=600$ K. %On the other hand for $\beta-$MoSi$_2$ the optimum VBA is observed for $T<600$ K.

The explanation for the observed magnification in $PF$ and $S$ with volume, can be pinned down to the lowering in energy of the first CB, along multiple directions in its BZ observed in Fig.~\ref{Fig7}b. These changes in the bandstructure not only causes an increased DOS of carriers around the CBM, but also an increased number of electron pockets. The Fermi surface demonstrating the electronic contributions at $V_{\mathrm{opt}}$ under doped scenario is shown in Fig.~\ref{MoSi2_fermi}. Both the above discussed factors improve $PF$ and $S$ for n-doped MoSi$_2$ with volume enlargement. Consequently at a doping of $n=2 \times 10^{20}$ cm$^{-3}$, we obtain the observe the largest VBA effect in this work of $zT^{\mathrm{\beta \; MoSi_2}}_{\mathrm{opt}}=1.07$ and $zT_{\mathrm{opt}}/zT_0=3.84$ (at $T=450$~K) for $n$ doping scenario. The same values at $T=600$~K are tabulated in Table.~\ref{Table1}.

%The possibilities to obtain the desired volume changes by alloying are isostructural hex MoGe$_2$ and MoSn$_2$ compounds. While hex MoGe$_2$ is on the respective Mo-Ge convex-hull, MoSn$_2$ is $\approx180$ meV from its Mo-Sn convex-hull. We will discuss more on the stabilities of the alloy in the next section. 

\section{Conclusion}

In this paper a computational HT-scheme to identify compounds where the thermoelectric properties can be optimized by alloying is presented. %of existing materials and also identifies several of these materials as promising candidates for thermoelectric applications. Furthermore, almost all of these selected materials will result in cost-effective thermoelectrics as they are comprised of elements that are readily available in the earth's crust. 

We confirm that Mg$_2$Si and Mg$_2$Ge exhibit large enhancement of their thermoelectric properties with volume. We report for the first time that Ca$_2$Si and Ca$_2$Ge, hexagonal MoSi$_2$ and Ca$_{9}$Ge$_{5}$ could exhibit increased thermoelectric properties due to a volumetric band alignment. In the cases of Ca$_2$Si and Ca$_2$Ge the solid-solutions with Sn can be obtained at the expense of a negligible amount of mixing energy and therefore the volume changes can be obtained by alloying. Among the remaining candidates, the volume increase for the VBA effect is thermodynamically difficult to achieve by alloying. However, we have obtained that $\beta-$MoSi$_2$ exhibits a significant increase in its thermoelectric properties due VBA, by the virtue of favorable changes to its bandstructure with volume. Thus, in this case even a small volume change due to alloying or thermal expansion may enhance its thermoelectric properties. Finally, we have established that $\Delta E_{\mathrm{h}}^{\mathrm{Sn}}$ can be a reliable descriptor to provide initial information on the stability of the corresponding alloys.

We have focused on systems where the electronic PF can be optimized with controlled volume changes. Alloying will also decrease the lattice part of the thermal conductivity, which will further improve the thermoelectric performance of the candidates. 

%($\kappa_{\mathrm{L}}$) through different mechanisms. Heat carrying phonons are increasingly scattered in a multi-component system. Furthermore, there are increased phonon-defect scattering in an alloy which will also reduce $\kappa_{\mathrm{L}}$. Since estimating the magnitude of $\kappa_{\mathrm{L}}$ can be computationally expensive \cite{Katre_JAP2015,Cormac_PRB2014}, we use a simplified expression for $\kappa_{\mathrm{min}}$ that captures the effect of alloying \cite{Levi2004}. We obtain that $\kappa_{\mathrm{Mg_2Si_{3/4}Sn_{1/4}}}=0.76\;\kappa_{\mathrm{Mg_2Si}}$ and $\kappa_{\mathrm{Ca_2Si_{3/4}Sn_{1/4}}}=0.77\;\kappa_{\mathrm{Ca_2Si}}$, with this simple model. Therefore, for the VBA candidates discussed in this work, we would expect a lower thermal conductivity and thus improved $zT$s.}

%Alloying also for lowering kappa

%The primary reason behind the volumetric enhancement can be entirely pinned on the change in the bandstructure, i.e. the situation resulting in lowering of a band near the CBM, thereby causing an extra contribution to the polupation of free electrons contributing to the thermoelectric properties.
\section*{Acknowledgments}

The authors would like to acknowledge funding from the Deutsche Forschungsgemeinschaft (DFG) grant numbers: MA 5487/1-1 and MA 5487/4-1.

%%%%%%%%%%%%%%%%%%%%%%%%%%%%%%%%%%%%%%%%%%%%%%%%%%%%%%

\bibliography{vol,te} %your .bib file
\bibliographystyle{rsc} %the RSC's .bst file

\end{document}